\documentclass[12pt]{article}
\pdfoutput=1
\usepackage{subfigure}
\usepackage{amssymb,amsmath}
\usepackage{graphicx}
\usepackage{color}
\usepackage[colorlinks=true
,urlcolor=blue
,citecolor=blue
,linkcolor=blue
,pagecolor=blue
,linktocpage=true
,pdfproducer=medialab
]{hyperref}
\usepackage[a4paper,width=15.2cm]{geometry}
\makeatletter \renewcommand{\@dotsep}{10000} \makeatother

\newcommand{\beq}{\begin{equation}}
\newcommand{\eeq}{\end{equation}}
\newcommand{\bea}{\begin{eqnarray}}
\newcommand{\eea}{\end{eqnarray}}

\begin{document}
%Remove date before submitting to arXiv
%\date{\today}

\begin{center}

 {\Large 125 GeV Higgs Boson from $t$-$b$-$\tau$ Yukawa Unification
 } \vspace{1cm}

{\large   Ilia Gogoladze\footnote{E-mail: ilia@bartol.udel.edu\\
\hspace*{0.5cm} On  leave of absence from: Andronikashvili Institute
of Physics, 0177 Tbilisi, Georgia.}, Qaisar Shafi\footnote{ E-mail:
shafi@bartol.udel.edu} and Cem Salih $\ddot{\rm U}$n \footnote{
E-mail: cemsalihun@bartol.udel.edu}} \vspace{.5cm}

{\baselineskip 20pt \it
Bartol Research Institute, Department of Physics and Astronomy, \\
University of Delaware, Newark, DE 19716, USA  } \vspace{.5cm}

\vspace{1.5cm}
 {\bf Abstract}
\end{center}

We identify a class of supersymmetric $SU(4)_c \times SU(2)_L \times SU(2)_R$ models
in which imposing essentially perfect $t$-$b$-$\tau$ Yukawa coupling
unification at $M_{\rm GUT}$ yields a mass close to 122-126 GeV for the lightest
CP-even (SM-like) Higgs boson. The squark and gluino masses in these
models exceed 3 TeV, but the stau and charginos in some cases can be
considerably lighter. We display some benchmark points corresponding to
neutralino-stau and bino-wino coannihilations as well as A-resonance.
The well-known MSSM parameter $ \tan \beta $ is around 46-52.

\newpage

%%%%%%%%%%%%%%%%%%%%%%%%%%%%%%%%%%%%%%%%%%%%%%%%%%%%%%%%%%%%
\renewcommand{\thefootnote}{\arabic{footnote}}
\setcounter{footnote}{0}

%%%%%%%%%%%%%%%%%%%%%%%%%%%%%%%%%%%%%%%%%%%%%%%%%%%%%%%%%%%%%

%\baselineskip 36pt
% Main body
%%%%%%%%%%%%%%%%%%%%%%%%%%
%\baselineskip 18pt
%%%%%%%%%%%%%%%%%%%%%%%%%%

\section{Introduction  \label{intro}}

In a recent paper, hereafter referred to as \cite{Gogoladze:2011aa}, we explored the LHC implications of a supersymmetric SO(10) model with third family Yukawa coupling unification  \cite{big-422,bigger-422,Baer:2008jn} and $ SO(10) $ compatible gaugino mass ratio at $ M_{\rm GUT} $ given by $M_3: M_2 : M_1= 2:-3:-1$ \cite{Martin:2009ad}. Here $M_3$, $M_2$ and $M_1$ respectively stand for  $SU(3)_c$, $SU(2)_L$ and $U(1)_Y$ soft supersymmetry breaking (SSB) gaugino masses.
It was also assumed in \cite{Gogoladze:2011aa} that the SSB $ {\rm mass}^{2} $ terms for the up and down type Higgs doublets of the minimal supersymmetric standard model (MSSM) are universal at $M_{GUT}$. It was shown in \cite{Gogoladze:2011aa} that $ t$-$b$-$\tau $ Yukawa coupling unification and neutralino dark matter abundance are readily compatible with each other in this $ SO(10) $ model, and a prediction of 122 - 124 GeV for the lightest CP-even  Higgs mass (with a theoretical uncertainty of $ \pm 3$ GeV \cite{Degrassi:2002fi}) was highlighted.

Spurred by these results, we investigate here $t$-$b$-$\tau$
Yukawa coupling unification  in the framework of
supersymmetric   $SU(4)_c \times SU(2)_L \times SU(2)_R$ 
(4-2-2, for short). The 4-2-2 structure \cite{pati} allows us to consider
non-universal gaugino masses, while preserving Yukawa coupling unification and universality of the up and down type SSB Higgs masses. Previously, $t$-$b$-$\tau$ Yukawa unification  in the framework of 4-2-2 was investigated in \cite{Gogoladze:2009ug,
Gogoladze:2009bn,Gogoladze:2010fu}, where non-universal up and down type SSB Higgs masses were assumed. An important conclusion reached in \cite{Gogoladze:2009ug,
Gogoladze:2009bn} is that with same sign non-universal gaugino soft mass terms and $ \mu > 0 $, where $ \mu $ denotes the coefficient of the bilinear Higgs mixing term, Yukawa unification in 4-2-2 is compatible with neutralino dark matter, with gluino co-annihilation \cite{Gogoladze:2009ug,
Gogoladze:2009bn, Baer:2009ff, Profumo:2004wk} being the unique dark matter scenario.

By considering opposite sign gauginos such that $\mu<0$, $M_2<0$ and $M_3>0$, it is shown in \cite{Gogoladze:2010fu} that Yukawa
coupling unification consistent with the experimental constraints
can be implemented in 4-2-2. Yukawa coupling unification in this case is achieved for $m_{16} \gtrsim 300\, {\rm
GeV}$, as opposed to $m_{16} \gtrsim 8\, {\rm TeV}$ for the case of
same sign gauginos. The finite corrections to the b-quark
mass play an important role here \cite{Gogoladze:2010fu}. With $M_2 <0$, $M_3>0$ and $\mu<0$ in 4-2-2, we can also obtain additional contributions with the correct sign for the muon anomalous magnetic moment \cite{Bennett:2006fi}. This enables us to simultaneously
satisfy the requirements of  $t$-$b$-$\tau$ Yukawa unification,
neutralino dark matter abundance and constraints from $(g-2)_\mu$, as well as a variety of
other bounds.

A crucial difference between this paper and previous studies of  4-2-2 with  $t$-$b$-$\tau$ Yukawa unification is the universality of the up and down type SSB Higgs masses assumed here. As we will see, the condition $m_{H_{u}}^{2}=m_{H_{d}}^{2}$ at $ M_{\rm GUT} $ significantly reduces the range of allowed masses for the lightest CP-even Higgs boson. With $m_{H_{u}}^{2} \neq m_{H_{d}}^{2}$ and imposing  $t$-$b$-$\tau$ Yukawa unification in 4-2-2, one finds $80\, \,\rm{GeV}\lesssim m_h \lesssim 128\,\, \rm{GeV}$. However, in our present model, this range shrinks to $122\,\,\rm{GeV} \lesssim m_h \lesssim 126\, \, \rm{GeV}$ \cite{Gogoladze:2010fu}, which is in good agreement with the recently reported evidence  from the ATLAS and CMS experiments~\cite{atlas_h,cms_h} for a Higgs boson of mass around 125 GeV.

The outline for the rest of the paper is as follows. In Section \ref{model} we briefly describe the SSB terms in the 4-2-2 model. In Section \ref{constraintsSection} we summarize the scanning procedure and the experimental constraints that we have employed. In Section \ref{results} we present our results, focusing in particular on the mass of the lightest (SM-like) CP-even Higgs boson. The squarks and gluino turn out to be relatively heavy, $ \gtrsim 3 $ TeV. However, the stau and chargino in some cases can be considerably lighter. The  table in this section highlights some benchmark points which can be tested at the LHC. Our conclusions are summarized in Section \ref{conclusions}.

%%%%%%%%%%%%%%%%%%%%%%%%%%%%%%%%%%%%%%%%%%%%

\section{Supersymmetric 4-2-2 and SSB Terms \label{model}}

In 4-2-2 the 16-plet of $SO(10)$ matter fields consists of
$\psi$ (4, 2, 1) and $\psi_c$ $(\bar{4}, 1, 2)$. The third family Yukawa coupling
$\psi_c \psi H$, where H(1,2,2) denotes the bi-doublet (1,2,2),  yields
the following relation at $M_{\rm GUT}$ \cite{big-422}:
\begin{align}
Y_t = Y_b = Y_{\tau} = Y_{\nu_{\tau}}. \label{f1}
\end{align}
Supplementing 4-2-2 with a discrete left-right (LR) symmetry
\cite{pati,lr} (more precisely C-parity) \cite{c-parity} reduces the
number of independent gauge couplings in 4-2-2 from three to two.
This is because C-parity imposes the gauge
coupling unification condition, $g_L=g_R$ at $M_{\rm GUT}$. We will
assume that due to C-parity the SSB
mass terms, induced at $M_{\rm GUT}$ through gravity mediated
supersymmetry breaking \cite{Chamseddine:1982jx}, are equal in magnitude for the  squarks
and sleptons of the three families. The tree level asymptotic SSB
gaugino masses, on the other hand, can be non-universal from the
following consideration. From C-parity, we can expect that the
gaugino masses at $M_{\rm GUT}$ associated with $SU(2)_L$ and
$SU(2)_R$ are the same ($M_2 \equiv M_2^R= M_2^L$). However, the
asymptotic $SU(4)_c$ and consequently $SU(3)_c$ SSB gaugino masses
can be different. With the hypercharge generator in 4-2-2 given by
$Y=\sqrt{2/5}~(B-L)+\sqrt{3/5}~I_{3R}$, where $B-L$ and $I_{3R}$ are
the diagonal generators of $SU(4)_c$ and $SU(2)_R$, we have the
following asymptotic relation among the three SSB gaugino masses:
\begin{align}
M_1=\frac{3}{5} M_2 + \frac{2}{5} M_3. \label{gaugino-condition}
\end{align}
The supersymmetric 4-2-2 model with C-parity that we consider thus has two
independent parameters ($M_2$ and $M_3$) in the gaugino sector.

As was pointed out in ref \cite{Gogoladze:2011aa}, with non-universal gaugino masses at $M_{GUT}$,  radiative electroweak symmetry breaking (REWSB) compatible with perfect or near-perfect $t$-$b$-$\tau$ Yukawa  unification can occur  even if we set $m_{H_{u}}=m_{H_{d}}$.  Note that this is quite difficult, if not impossible, to achieve if gaugino mass universality is imposed at $ M_{\rm GUT} $ \cite{Olechowski:1994gm}.

 The fundamental parameters of
the 4-2-2 model that we consider are as follows:

\begin{align}
m_{16}, m_{10}, M_2, M_3, A_0, \tan\beta, {\rm sign}(\mu).
\label{params}
\end{align}
Here $m_{16}$ is the universal SSB mass for MSSM sfermions, $m_{10}$  is the universal SSB mass for
MSSM Higgs doublets,
$A_0$ is the universal coupling of the SSB terms for the trilinear scalar interactions (with the
corresponding Yukawa coupling factored out), $\tan\beta$ is the
ratio of the vacuum expectation values  (VEVs) of the two MSSM Higgs
doublets, and the magnitude of $\mu$, but not its sign, is determined by the REWSB condition.
  Although not necessary, we will assume that the gauge coupling unification condition $g_3=g_1=g_2$
holds at $M_{\rm GUT}$ in 4-2-2. Such a scenario can arise,
for example, from a higher dimensional $SO(10)$ model
\cite{Hebecker:2001jb}, or from a $SU(8)$ model after suitable
compactification \cite{su8}.

%%%%%%%%%%%%%%%%%%%%%%%%%%%%%%%%%%%%%%%%%%%%%%%%%%%%%%%%%%%%%%

\section{Experimental Constraints, Scanning Procedure and Parameter Space\label{constraintsSection}}

We employ the ISAJET~7.80 package~\cite{ISAJET}  to perform random
scans over the fundamental parameter space. In this package, the weak scale values of gauge and third generation Yukawa
couplings are evolved to $M_{\rm GUT}$ via the MSSM renormalization
group equations (RGEs) in the $\overline{DR}$ regularization scheme.
We do not strictly enforce the unification condition $g_3=g_1=g_2$ at $M_{\rm
GUT}$, since a few percent deviation from unification can be
assigned to unknown GUT-scale threshold
corrections~\cite{Hisano:1992jj}.
The deviation between $g_1=g_2$ and $g_3$ at $M_{\rm GUT}$ is no
worse than $3-4\%$.
For simplicity,  we do not include the Dirac neutrino Yukawa coupling
in the RGEs, whose contribution is expected to be small.

The various boundary conditions are imposed at
$M_{\rm GUT}$ and all the SSB
parameters, along with the gauge and Yukawa couplings, are evolved
back to the weak scale $M_{\rm Z}$.
In the evolution of Yukawa couplings the SUSY threshold
corrections~\cite{Pierce:1996zz} are taken into account at the
common scale $M_{\rm SUSY}= \sqrt{m_{{\tilde t}_L}m_{{\tilde t}_R}}$, 
where $m_{{\tilde t}_L}$ and $m_{{\tilde t}_R}$
denote the masses of the third generation left and right-handed stop quarks.
The entire parameter set is iteratively run between $M_{\rm Z}$ and $M_{\rm
GUT}$ using the full 2-loop RGEs until a stable solution is
obtained. To better account for leading-log corrections, one-loop
step-beta functions are adopted for the gauge and Yukawa couplings, and
the SSB parameters $m_i$ are extracted from RGEs at multiple scales
$m_i=m_i(m_i)$. The RGE-improved 1-loop effective potential is
minimized at $M_{\rm SUSY}$, which effectively
accounts for the leading 2-loop corrections. Full 1-loop radiative
corrections are incorporated for all sparticle masses.

The approximate error of  $\pm 3$ GeV in the ISAJET estimation of the lightest CP-even Higgs boson mass
largely arises from theoretical  uncertainties \cite{Degrassi:2002fi}  in the calculation   and
to a lesser extent from experimental uncertainties.

An important constraint on the parameter space arises from limits on the cosmological abundance of stable charged
particles  \cite{Nakamura:2010zzi}. This excludes regions in the parameter space
where  charged SUSY particles  become
the lightest supersymmetric particle (LSP). We accept only those
solutions for which one of the neutralinos is the LSP and saturates
the WMAP  bound on relic dark matter abundance.

We have performed random scans for the following parameter range:
\begin{align}
0\leq  m_{16}  \leq 6\, \rm{TeV} \nonumber \\
0\leq   m_{10} \leq 5\, \rm{TeV} \nonumber \\
0 \leq M_{3}  \leq 5 \, \rm{TeV} \nonumber \\
-5 \rm{TeV} \leq M_{2}  \leq 0  \nonumber \\
35\leq \tan\beta \leq 55 \nonumber \\
-3\leq A_{0}/m_{16} \leq 3\nonumber\\
\mu < 0
 \label{parameterRange}
\end{align}
We set    $m_t = 173.3\, {\rm GeV}$  \cite{:1900yx}, and we show that our results are not
too sensitive to one or two sigma variation from the central value of $m_t$  \cite{Gogoladze:2011db}.
Note that $m_b(m_Z)=2.83$ GeV, which is hard-coded into ISAJET.

In order to obtain the correct sign for the desired contribution to $(g-2)_{\mu}$, we will focus here on the case  $\mu<0$ and $M_2<0$.

In scanning the parameter space, we employ the Metropolis-Hastings
algorithm as described in \cite{Belanger:2009ti}. The data points
collected all satisfy
the requirement of REWSB,
with the neutralino in each case being the LSP. After collecting the data, we impose
the mass bounds on all the particles \cite{Nakamura:2010zzi} and use the
IsaTools package~\cite{Baer:2002fv}
to implement the various phenomenological constraints. We successively apply the following experimental constraints on the data that
we acquire from ISAJET:
\begin{table}[h]\centering
\begin{tabular}{rlc}
$m_h~{\rm (lightest~Higgs~mass)} $&$ \geq\, 114.4~{\rm GeV}$          &  \cite{Schael:2006cr} \\
$BR(B_s \rightarrow \mu^+ \mu^-) $&$ <\, 4.5 \times 10^{-9}$        &   \cite{:2007kv}      \\
$2.85 \times 10^{-4} \leq BR(b \rightarrow s \gamma) $&$ \leq\, 4.24 \times 10^{-4} \;
 (2\sigma)$ &   \cite{Barberio:2008fa}  \\
$0.15 \leq \frac{BR(B_u\rightarrow
\tau \nu_{\tau})_{\rm MSSM}}{BR(B_u\rightarrow \tau \nu_{\tau})_{\rm SM}}$&$ \leq\, 2.41 \;
(3\sigma)$ &   \cite{Barberio:2008fa}  \\
$\Omega_{\rm CDM}h^2 $&$ =\, 0.1123 \pm 0.0035 \;(5\sigma)$ &
\cite{Komatsu:2008hk} \\ $ 0 \leq \Delta(g-2)_{\mu}/2 $ & $ \leq 55.6 \times 10^{-10} $ & \cite{Bennett:2006fi}
\end{tabular}\label{table}
\end{table}

%%%%%%%%%%%%%%%%%%%%%%%%%%%%%%%%%%%%%%%%%%%%%%%%%%%%%%%%%%%%%%%%%%%%%%%%%

\section{Sparticle Spectroscopy \label{results}}

\begin{figure}[t!]
\centering
\includegraphics[width=14.7cm]{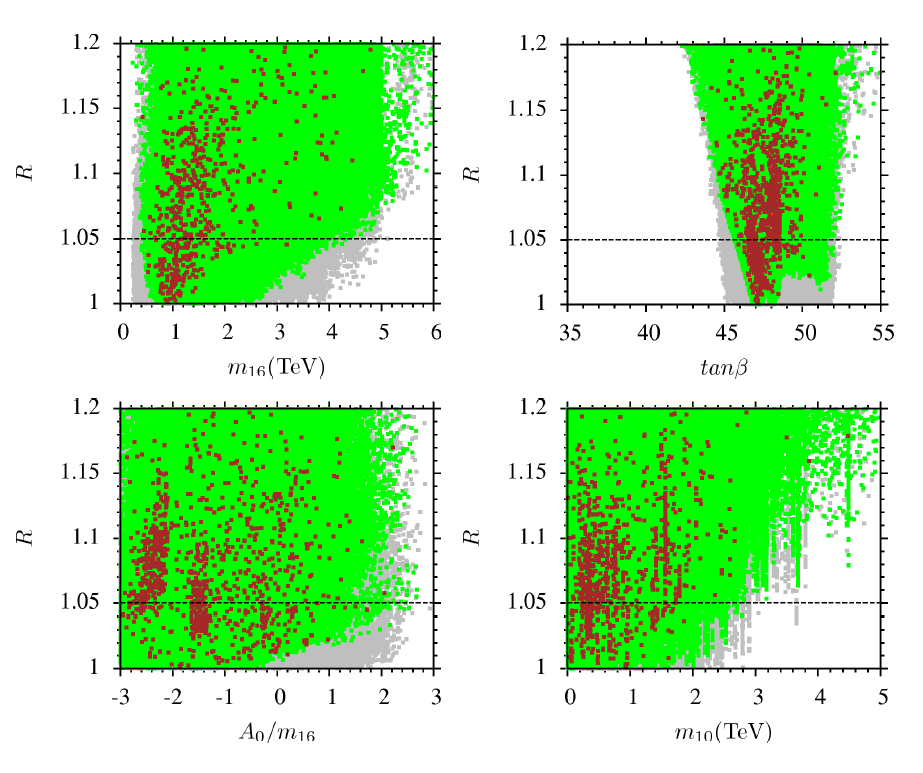}
%\vspace{1.5cm}
\caption{Plots in $ R-m_{16} $, $ R-\tan\beta $, $ R-A_{0}/m_{16} $ and $ R - m_{10}$ planes. Gray points are consistent with REWSB and neutralino LSP. Green points satisfy particle mass bounds and constraints from $BR(B_s\rightarrow \mu^{+} \mu^{-})$, $BR(b\rightarrow s \gamma)$ and $BR(B_u\rightarrow \tau \nu_\tau)$. In addition, we require that green points do no worse than the SM in terms of $ (g-2)_{\mu} $. Brown points belong to a subset of green points and satisfy the WMAP bounds on neutralino dark matter abundance.}
\label{fund-1}
\end{figure}

\begin{figure}[t!]
\centering
\includegraphics[width=14.7cm]{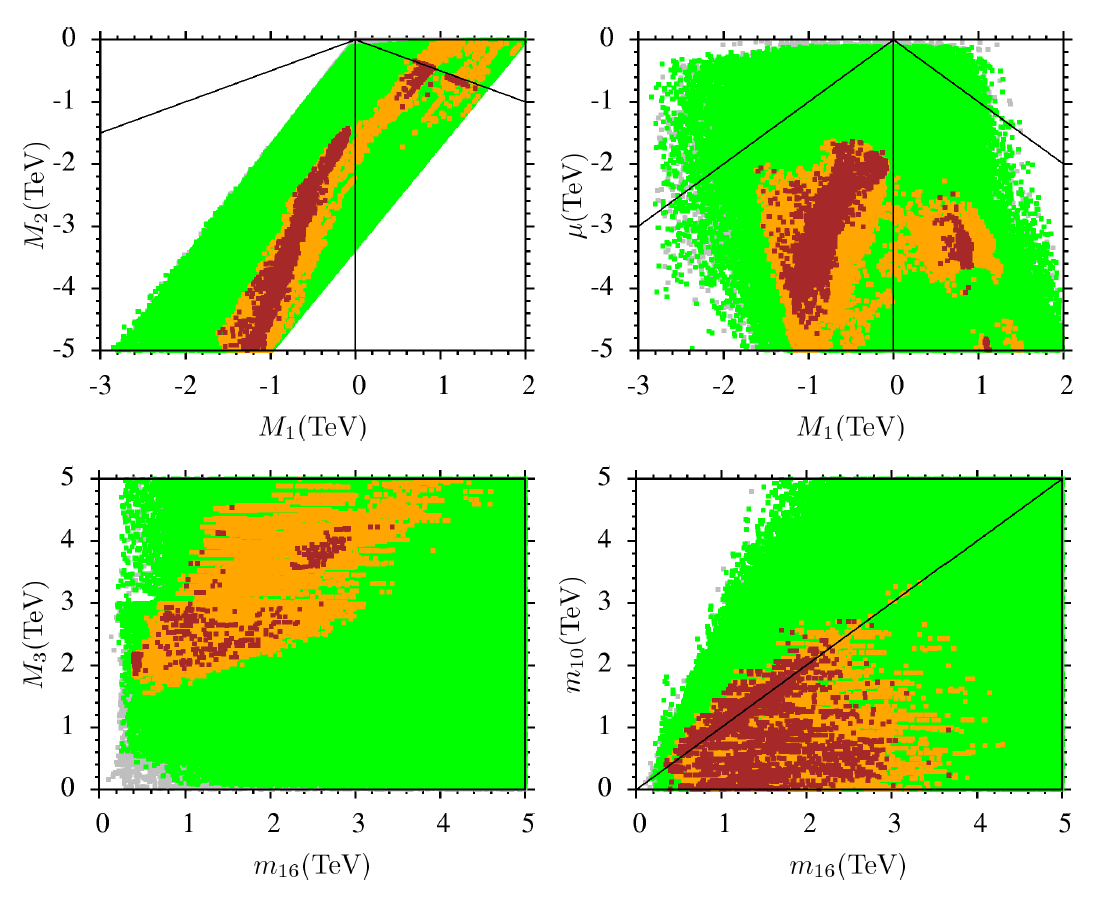}
\caption{Plots in $ M_2-M_{1} $, $ \mu-M_{1} $, $ M_{3}-m_{16} $ and $ m_{10}-m_{16} $ planes. The color coding is the same as in Figure \ref{fund-1}. In addition, we have used  yellow color to denote a subset of the green points, that have Yukawa unification better than $ 5\% $.
Brown points belong to a subset of yellow points and  satisfy the WMAP bounds on neutralino dark matter abundance.}
\label{fund-2}
\end{figure}

\begin{figure}[t!]
\centering
\includegraphics[width=15.2cm]{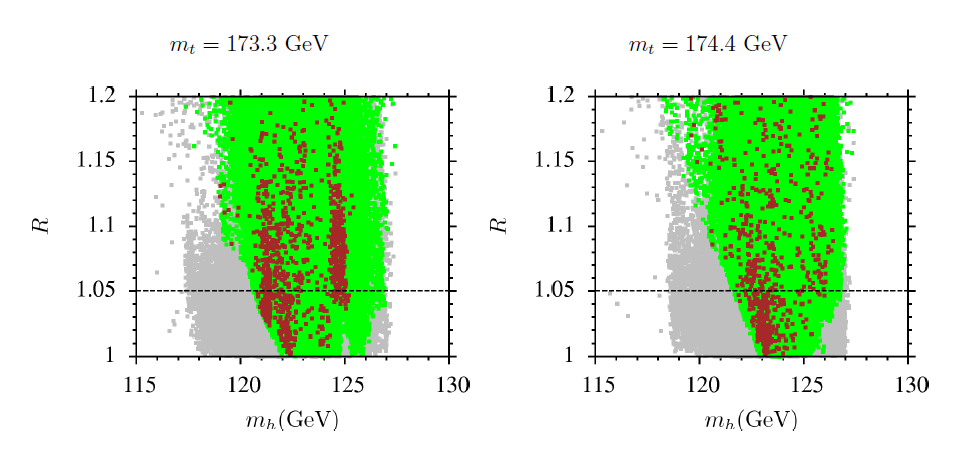}
\caption{Plots in the $ R-m_{h} $ plane. The color coding is the same as  in Fig \ref{fund-1}. In the left panel, the top mass is set equal to its current central value ($ m_{t}=173.3~{\rm GeV} $ \cite{:1900yx}), while 1$ \sigma $ deviation in the top mass ($ m_{t}=174.4~{\rm GeV} $) is considered in the right panel to investigate the effect on the CP-even light higgs mass.}
\label{fund-3}
\end{figure}

\begin{figure}[t!]
\centering
\includegraphics[width=12.3cm]{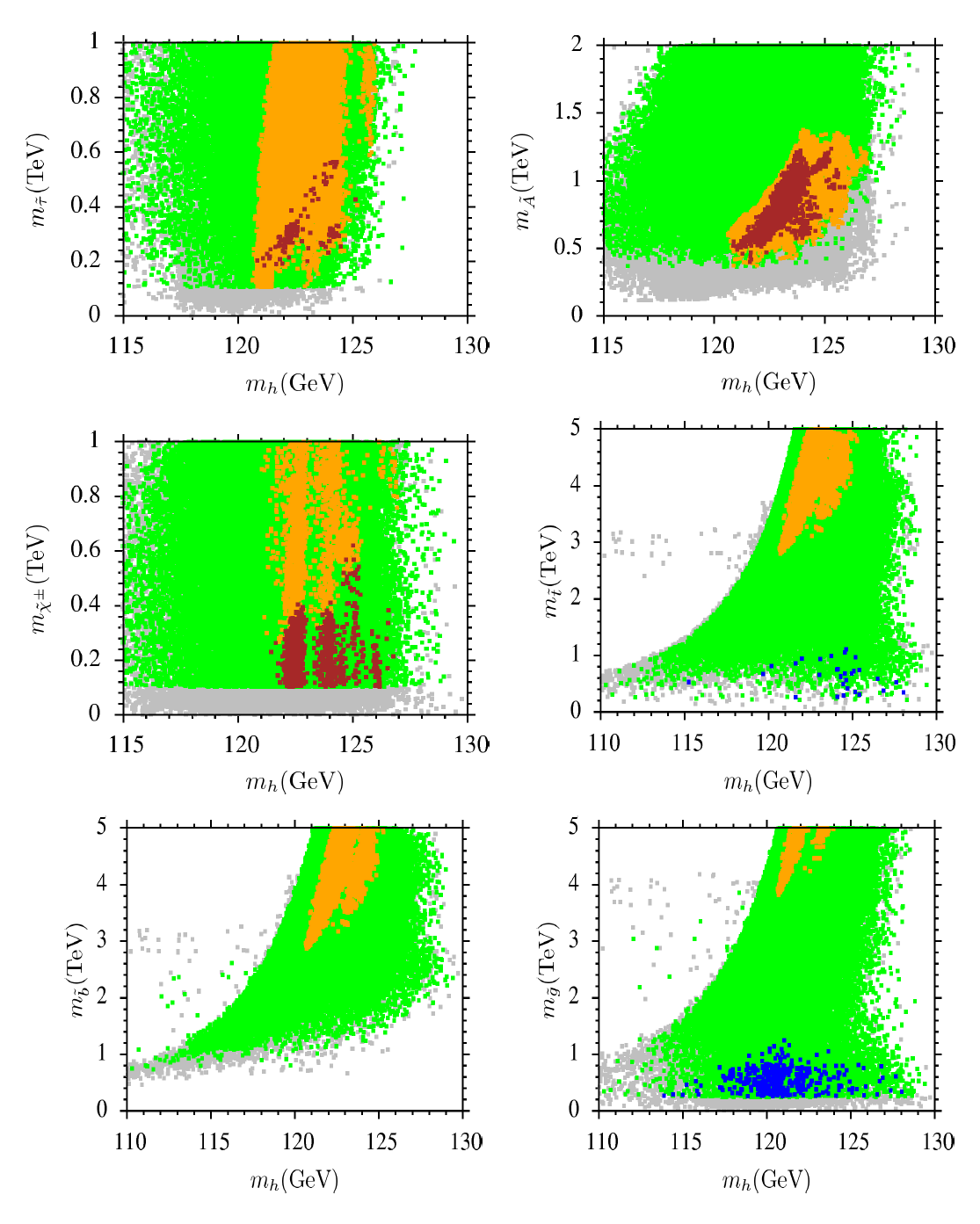}
\caption{Plots in the $ m_{\tilde{\tau}}-m_{h} $, $ m_{A}-m_{h} $, $ m_{\tilde{\chi}^{\pm}}-m_{h} $, $ m_{\tilde{t}}-m_{h} $, $ m_{\tilde{b}}-m_{h} $ and $ m_{\tilde{g}}-m_{h} $ planes. The color coding is the same as in Figure 1. The brown points in the $ m_{\tilde{\tau}}-m_{h} $ plane represent stau-neutralino coannihilation, A-funnel region  in the $ m_{A}-m_{h} $ panel, bino-wino dark mater  and bino-wino coannihilation  in  the $ m_{\tilde{\chi}^{\pm}}-m_{h} $ plane.
In the $ m_{\tilde{t}}-m_{h} $, $ m_{\tilde{b}}-m_{h} $ and $ m_{\tilde{g}}-m_{h} $ planes, the blue points form a subset of the green points and do not satisfy Yukawa unification. Blue points represent stop-neutralino coannihilation in the $ m_{\tilde{t}}-m_{h} $ plane, and gluino-neutralino coannihilation in the $ m_{\tilde{g}}-m_{h} $ plane. }
\label{fund-4}
\end{figure}

In order to quantify Yukawa coupling unification, we define the quantity $ R $ as,

\begin{align}
R=\frac{ {\rm max}(y_t,y_b,y_{\tau})} { {\rm min} (y_t,y_b,y_{\tau})}
\end{align}
In Figure \ref{fund-1} we show the results in the $ R-m_{16} $, $ R-{A_{0}/m_{16}} $ and $ R-m_{10} $ planes. The gray points are consistent with REWSB and neutralino LSP. The green points satisfy particle mass bounds and constraints from $ BR(B_{s}\rightarrow \mu^{+}\mu^{-}) $, $ BR(b\rightarrow s \gamma) $ and $ BR(B_{u}\rightarrow \tau\nu_{\tau}) $. In addition, the green points do no worse than the SM in terms of $ (g-2)_{\mu} $. The brown points belong to a subset of the green points and satisfy the WMAP bound on neutralino dark matter abundance.

In the $ R-m_{16} $ plane of Figure \ref{fund-1}, we can see  realization of  perfect Yukawa unification consistent with all constraints mentioned in Section \ref{constraintsSection}. This is possible because we can implement Yukawa unification for relatively small values of $ m_{16} $ ($ \sim 1 {\rm TeV} $).
A similar conclusion was reached in ref. \cite{Gogoladze:2010fu} but with relatively lighter values for $ m_{16} $. This difference is related to the additional condition $m_{H_{u}}=m_{H_{d}}$ which we have imposed in the current paper. 

Note that there is more than an order of magnitude reduction in the $ m_{16} $ values required for Yukawa unification, as compared with the case $\mu > 0 $ and universal gaugino masses \cite{Baer:2008jn, Gogoladze:2009ug}. There is upper bound for $ m_{16} $  once Yukawa unification is imposed.
  Indeed, one predicts $ \tan \beta \approx 47 $ for perfect Yukawa unification. In $ R-A_{0}/m_{16} $ plane, we see that perfect Yukawa unification can occur for  $ A_{0} < 0 $,   in contrast to the SO(10) case with universal gaugino mass condition where perfect Yukawa unification takes place for $A_0/m_{16} \sim 2.6$.
From the $ R-m_{10} $ plane we see that perfect Yukawa unification also puts an upper bound on $m_{10}$.

In Figure \ref{fund-2} we show the results in the  $ M_2-M_{1} $, $ \mu-M_{1} $, $ M_{3}-m_{16} $ and $ m_{10}-m_{16} $ planes. The color coding is the same as in Figure \ref{fund-1}. In addition, we have used  yellow color  for the subset of green points satisfying Yukawa unification to better than $ 5\% $.
The brown points belong to a subset of green points which satisfy the WMAP bounds on neutralino dark matter abundance.

From the $ M_2-M_{1} $ plane, we see that some of the brown points lie near the line corresponding to $M_1/M_2\approx -2$, which yields at low sale $M_{\tilde B}\simeq M_{\tilde W}$, the condition for bino-wino coannihilation \cite{Baer:2005jq} scenario. At the same time there is no colored point near the line corresponding to $M_1/M_2\approx 2$, which means that in our model we do not have bino-wino mixed dark matter   \cite{Baer:2006dz}.

In the $ \mu-M_{1} $ plane we have shown lines corresponding to $\mu/M_1=1$ and $\mu/M_1=-1$. The yellow points describing $ 5\% $ or better Yukawa unification are far from these lines, which means that we do not have bino-higgsino mixed dark matter in this model. Accordingly, as shown in ref \cite{Gogoladze:2010ch}, it will be difficult to find dark matter predicted in this model in direct and indirect searches. 

The $M_3-m_{16}$ panel indicates that the minimal value of $M_3$ which corresponds to $5\%$ or better Yukawa unification (yellow points) is above 1.5 TeV, which means that the colored particle are much heavier than 1 TeV. In particular, the minimum gluino mass is close to 3 TeV.

From the $m_{10}-m_{16}$ plane we see that one can reduce number of independent parameters, assuming $m_{16}=m_{10}$, and still retain decent Yukawa unification with the correct relic abundances.

In Figure  \ref{fund-3}, we present results in the $ R-m_{h} $ plane for $ m_{t}=173.3 $ GeV and $ m_{t}=174.4 $ GeV. The color coding is the same as the one used in Figure \ref{fund-1}. In the left panel, the top quark mass is set equal to its central value ($ m_{t} = 173.3~{\rm GeV}$  \cite{:1900yx}), and it shows that we can find solutions with $ m_{h} \sim 125~{\rm GeV}$ consistent with the constraints mentioned in Section \ref{constraintsSection} as well as $ t $-$ b $-$ \tau $ Yukawa unification.  It is interesting to note that the Higgs mass interval is significantly reduced compared with the 4-2-2 model with $m_{H_{u}}\neq m_{H_{d}}$ where it lies in the interval $80~ {\rm GeV} < m_h < 129$ GeV, compatible with perfect  Yukawa unification condition. In the current model, as seen from Figure 3, the Higgs mass lies in the range  $118~ {\rm GeV} < m_h < 127$ GeV (gray points). This is further reduced once collider bounds and precise Yukawa unification are imposed (green points) and the range becomes $122\,\, {\rm GeV} < m_h < 126$ GeV, which is in very good agreement with the recent ATLAS and CMS measurements \cite{atlas_h,cms_h}. 

As far as neutralino dark matter abundance compatible with the WMAP bound is concerned, in Figure \ref{fund-4} we present results in the $ m_{\tilde{\tau}}-m_{h} $, $ m_{A}-m_{h} $, $ m_{\tilde{\chi}^{\pm}}-m_{h} $, $ m_{\tilde{t}}-m_{h} $, $ m_{\tilde{b}}-m_{h} $ and $ m_{\tilde{g}}-m_{h} $ planes. The color coding is the same as in Figure 1. The brown points in the $ m_{\tilde{\tau}}-m_{h} $ plane represent only stau-neutralino coannihilation, they also represent A-funnel solution  in the $ m_{A}-m_{h} $ panel, as well as bino-wino dark matter  and bino-wino coannihilation  in the $ m_{\tilde{\chi}^{\pm}}-m_{h} $ plane.
In the $ m_{\tilde{t}}-m_{h} $, $ m_{\tilde{b}}-m_{h} $ and $ m_{\tilde{g}}-m_{h} $ planes, the blue points are a subset of the green points and do not satisfy Yukawa unification.

If the preliminary evidence for a Higgs with  $m_h=125$ GeV from ATLAS/CMS is verified, we see from the $ m_{\tilde{\tau}}-m_{h} $ panel solutions with $ 200~{\rm GeV} \lesssim m_{\tilde{\tau}} \lesssim 600~{\rm GeV} $,  linked to stau-neutralino coannihilation. Similarly, the $ m_{A}-m_{h} $ plane shows $ A $-funnel solutions with $ 500~{\rm GeV}\lesssim m_{A} \lesssim 1200~{\rm GeV} $. The $ m_{\tilde{\chi}^{\pm}}-m_{h} $ panel shows that we can have bino-wino dark matter, and bino-wino coannihilation is realized in nature with
 $100~ {\rm GeV} \lesssim  m_{\tilde{\chi}^{\pm}} \lesssim 600$ GeV.
We did not find solutions with neutralino-stop coannihilation compatible with $5\%$ or better Yukawa unification. However, such solutions appear if the Yukawa unification condition is relaxed (blue points). In the  $m_{\tilde{b}}-m_{h} $ plane we do not see either brown or blue points which shows that in the 4-2-2 model it is not easy to find neutralino-sbottom coannihilation, a scenario which is easily realized in SU(5) \cite{Gogoladze:2011ug}. We also did not find neutralino-gluino coannihilation compatible with Yukawa unification (no brown points in the $ m_{\tilde{g}}-m_{h} $ plane), but there are blue points which show that we can have neutralino-gluino coannihilation for NLSP gluino up to 1.5 TeV mass, if we ignore Yukawa unification.

Finally, in Table \ref{tab1} we present three benchmark points which satisfy the various constraints mentioned in Section \ref{constraintsSection}. Point 1 displays a solution corresponding to stau-neutralino coannihilation, point 2 depicts an A-funnel solution, point 3 stands for a solution with bino-wino coannihilation, and point 4 corresponds to a solution with $ m_{16}=m_{10} $. Each point in Table \ref{tab1} has the CP-even lightest Higgs boson mass of about 125 GeV.

\begin{table}[t!]\vspace{1.5cm}
\centering
\begin{tabular}{|c|cccc|}
\hline
\hline
                 & Point 1 & Point 2 & Point 3 & Point 4\\
\hline
$m_{16}$         & 1729  & 2777  & 2406 & \textbf{2016}\\
$ M_{1} $        & -809 & 1262  & \textbf{1177} & -1046 \\
$M_{2} $         & -4180  & -685  & \textbf{-633} & -4867\\
$M_{3} $         & 4247  & 4183  & 3892 & 4686\\
$m_{10}$         & 965  & 640  & 227 & \textbf{2016}\\
$\tan\beta$      & 49.2  & 48.6 & 48.2 & 48.2\\
$A_0/m_0$        & 1.04  & -2.74 & -2.82 & -2.81\\ 
$m_t$            & 173.3   & 173.3 & 173.3 & 173.3\\
\hline
$\mu$            & -3762  & -5765  & -5305 & -4539\\
$\Delta(g-2)_{\mu}$  & $0.62\times 10^{-10} $ & $0.50\times 10^{-10}$  & $0.59\times 10^{-10} $ & $0.43\times 10^{-10}$\\

\hline
$m_h$            & \textbf{125}  & \textbf{125} & \textbf{125} & \textbf{125}\\
$m_H$            & 949  & 1160  & 1197 & 1048\\
$m_A$            & 943  & \textbf{1152}  & 1190 & \textbf{1042}\\
$m_{H^{\pm}}$    & 954  & 1164    & 1201 & 1053\\

\hline
$m_{\tilde{\chi}^0_{1,2}}$
                 & \textbf{405}, 3608 & \textbf{556}, 663  & \textbf{515, 614} & \textbf{508}, 4188\\
$m_{\tilde{\chi}^0_{3,4}}$
                 & 3769, 3795 & 5739, 5740  & 5283, 5283 & 4540, 4552\\

$m_{\tilde{\chi}^{\pm}_{1,2}}$
                 & 3605, 3812 & 666, 5738  & 616, 5282 & 4193, 4555\\
$m_{\tilde{g}}$  & 8496 & 8455  & 7891 & 9340\\

\hline $m_{ \tilde{u}_{L,R}}$
                 & 7828, 7386  & 7691, 7705  & 7128, 7140 & 8674, 8138\\
$m_{\tilde{t}_{1,2}}$
                 & 6265, 6943 & 5978, 6209  & 5579, 5800 & 6541, 7387\\
\hline $m_{ \tilde{d}_{L,R}}$
                 & 7828, 7388  & 7692, 7706  & 7128, 7141 & 8675, 8139\\
$m_{\tilde{b}_{1,2}}$
                 & 6430, 6903  & 6128, 6225  & 5716, 5813 & 6743, 7346\\
\hline
$m_{\tilde{\nu}_{1}}$
                 & 3147  & 2807  & 2435 & 3650\\
$m_{\tilde{\nu}_{3}}$
                 &  2909 & 2344  & 2037 & 3376\\
\hline
$m_{ \tilde{e}_{L,R}}$
                & 3155, 1743  & 2810, 2810  & 2439, 2439 & 3658, 2040\\
$m_{\tilde{\tau}_{1,2}}$
                & \textbf{426}, 2925  & 1617, 2330  & 1388, 2028 & 710, 3402\\
\hline

$\sigma_{SI}({\rm pb})$
                & $0.67\times 10^{-11}$  & $0.11\times 10^{-11}$  & $0.11\times 10^{-11} $ & $0.36\times 10^{-11}$\\

$\sigma_{SD}({\rm pb})$
                & $0.46\times 10^{-9}$  & $0.30\times 10^{-10}$  & $0.45\times 10^{-10} $ & $0.20\times 10^{-9}$\\

$\Omega_{CDM}h^{2}$
                &  \textbf{0.26} & \textbf{0.11}  & \textbf{0.11} & \textbf{1.22}\\
\hline

$R$     &\textbf{1.03}  & \textbf{1.02}  & \textbf{1.04} & \textbf{1.03}\\

\hline
\hline
\end{tabular}
\caption{Benchmark points for the 4-2-2 model with $ \mu < 0 $. Point 1 shows stau-coannihilation, point 2 represents the $ A $-resonance solution, point 3 depicts bino-wino coannihilation, and point 4 displays a solution with $ m_{16}=m_{10} $.}
\label{tab1}
\end{table}

%%%%%%%%%%%%%%%%%%%%%%%%%%%%%%%%%%%%%%%%%%%%%%

\section{Conclusion \label{conclusions}}
We have reconsidered $t$-$b$-$\tau$ Yukawa unification in supersymmetric $ SU(4)_{c}\times SU(2)_{L}\times SU(2)_{R} $ within a slightly revised framework in this paper. The main difference from most previous investigations stems from the assumptions we make related to the soft supersymmetry breaking parameters.  We set the masses of the two MSSM Higgs doublets to be equal at $M_{\rm GUT}$.
The ramifications of these slightly different assumptions for TeV scale physics turn out to be quite startling, with the low energy predictions being quite different from previous studies. In particular, in the present framework with $ \mu $ and $ M_{2} $ both negative, demanding perfect or near-perfect Yukawa unification yields a Higgs boson mass of 122 - 125 GeV (with an uncertainty of $ \pm 3 $ GeV).

The solutions, obtained using the ISAJET  software, are compatible with all experimental observations, as well as the WMAP dark matter constraint. The masses of the gluino and first two family squarks are found to lie in the 2.7 - 5 TeV range, while the lightest stop (top squark) weighs at least 2 TeV or so. The MSSM parameter $\tan\beta$ is around 46 - 52.

\newpage

%%%%%%%%%%%%%%%%%%%%%%%%%%%%%%%%%

\end{document}